\documentclass[12pt, a4paper]{article}
\pdfoutput=1

\usepackage{amsmath}
\usepackage{amsfonts}
\usepackage{amssymb}
\usepackage{graphicx, rotating}
\usepackage{epsfig}
\usepackage{latexsym}
\usepackage{graphicx}
\usepackage{color}
\usepackage{amsmath,bm,amssymb}
\usepackage{cite}
\usepackage{slashed}
\usepackage{hyperref}
\hypersetup{colorlinks, citecolor=bluscuro, linkcolor=black, urlcolor=bluscuro}
\definecolor{rossos}{cmyk}{0,1,1,0.55}
\definecolor{bluscuro}{rgb}{0.15, 0.2, .85}
\definecolor{bluchiaro}{cmyk}{1,.3,0.,0.1}

\newcommand{\be}{\begin{equation}}
\newcommand{\ee}{\end{equation}}
\newcommand{\bea}{\begin{eqnarray}}
\newcommand{\eea}{\end{eqnarray}}

\def\Gb{\overline G}

\newcommand{\arXiv}[2]{\href{http://arxiv.org/pdf/#1}{{\tt [#2/#1]}}}
\newcommand{\arXivold}[1]{\href{http://arxiv.org/pdf/#1}{{\tt [#1]}}}

\def\bma#1{\mbox{\boldmath{$#1$}}}

\begin{document}
\allowdisplaybreaks
\begin{titlepage}
\begin{flushright}
DESY 17-226
\end{flushright}
\vspace{.3in}

\vspace{1cm}
\begin{center}
{\Large\bf\color{black}
Resummation of Goldstone Infrared Divergences: \\[0.5cm] 
A Proof to All Orders}\\
\bigskip\color{black}
\vspace{1cm}{
{\large J.R.~Espinosa$^{a,b}$, T.~Konstandin$^c$}
\vspace{0.3cm}
} \\[7mm]
{\it {$^a$\, Institut de F\'isica d'Altes Energies (IFAE), The Barcelona Institute of Science and Technology (BIST), Campus UAB, 08193 Bellaterra (Barcelona), Spain}}\\
{\it $^b$ {ICREA,  Instituci\'o Catalana de Recerca i Estudis Avan\c{c}ats,\\ Pg. Llu\'is Companys 23, 08010 Barcelona, Spain}}\\
{\it $^c$ {DESY, Notkestr. 85, 22607 Hamburg, Germany}}
\end{center}
\bigskip

\vspace{.4cm}

\begin{abstract}
The perturbative effective potential calculated in Landau gauge suffers 
from infrared problems due to Goldstone boson loops. These divergences are spurious and can be removed by a resummation procedure that amounts to a shift of the mass of soft Goldstones. We prove this to all loops using an effective theory approach, providing a compact recipe for the shift of the Goldstone mass that relies on the use of the method of regions to split soft and hard Goldstone contributions.
\end{abstract}
\bigskip

\end{titlepage}

\section{Introduction \label{sec:intro}} 

The effective potential is widely used in many areas of particle physics and cosmology. Among other applications, it is the central tool to study
symmetry breaking in many contexts, phase transitions at finite temperature~\cite{VT}, the slow-roll evolution of the inflaton field \cite{Vinf}, etc. The radiatively corrected potential \cite{Vrad}
has been used to study the radiative breaking of symmetries \cite{CW} (leading to dimensional transmutation) and it is as well an efficient way to calculate radiative corrections to the Higgs mass in many beyond the Standard Model (BSM) scenarios (see {\it e.g.} \cite{SMH,MSSMH}).
In the context of the Standard Model (SM), the effective Higgs potential describes the spontaneous breaking of the electroweak symmetry and the fate of the Standard Model vacuum at late times~\cite{STAB,stab}.
This SM potential has been known at two loops \cite{V2} since the early nineties. The two-loop potential for a generic renormalizable quantum field  theory was obtained in \cite{V2M} and the three-loop corrections have been  obtained quite recently in a tour-de-force calculation by Steve P. Martin in \cite{V3}. 

The SM effective potential was calculated in the above papers in Landau gauge using the minimal subtraction scheme ($\overline{\mathrm{MS}}$) and dimensional regularization. This effective potential suffers from infrared (IR) problems due to loops involving Goldstone bosons, which in this scheme are massless close to the vacuum.
More specifically, writing the tree-level potential as
\be
V_0(\phi) =-\frac12 m^2 \phi^2 +\frac14 \lambda \phi^4\ ,
\label{V0}
\ee
where $\phi$ is the real part of the neutral component of the Higgs doublet, $\phi \equiv \sqrt{2}\,\mathrm{Re}(H^0)$, the tree-level
Goldstone mass is
\be
G\equiv \frac{1}{\phi}\frac{\partial V_0}{\partial\phi} = -m^2+\lambda \phi^2\ .
\ee
Let us write the radiatively corrected effective potential  as
\be
V=V_0 + \kappa V_1  + \kappa^2 V_2 + ...\ ,
\ee
where we have pulled out powers of $\kappa=1/(16\pi^2)$ to indicate the loop-order of each correction. This potential, as well as its derivatives, is IR divergent for $G\rightarrow 0$. Calling $X$ a generic 
squared mass that does not vanish for $G\rightarrow 0$ (like $T=h_t^2\phi^2/2$ for the top quark), the IR divergence of $V$ first appears in $V_3$ through terms of the form $X^2\log G$, getting even worse at higher orders, with $V_{n\geq 4} \supset X^{n-1}/G^{n-3}$. The two-loop potential, $V_2$, is IR finite but contains terms $\sim X G\log G$ that make $V_2'\equiv \partial V_2/\partial \phi$ IR divergent. The Goldstone contribution to the one-loop potential, $V_1$, is of the form 
$\sim G^2\log G$ and leads to a divergence in $V_1''$.

Such IR divergences cause trouble when they appear in $V$, as they
would make it impossible to give a physical meaning to the potential,
or $V'$, as the determination of the minimum of the potential requires solving $V'=0$, but are not problematic in higher field derivatives. For instance, it is well known that the IR divergence in $V''$, used in calculating the Higgs mass via the effective potential method, is harmless. The on-shell Higgs self-energy, that enters the calculation of the physical pole Higgs mass, does not suffer from this divergence (see {\it e.g.} \cite{MH}), which only affects the self-energy at zero external momentum (the one $V''$ reproduces). From now on, with an slight abuse of terminology, we will refer to the IR divergences of the potential as those affecting $V$ and $V'$ only.
As we show in this paper, after appropriate resummation one ends up with a potential that has IR-finite $V$ and $V'$, but IR divergent higher derivatives. We consider such resummed potential as  IR-safe.

This Goldstone IR problem was first noticed in~\cite{IR} and 
emphasized more recently in \cite{MartinCatast}. Shortly afterwards, it was realized in~\cite{MartinIR} and~\cite{USIR} that this issue can be resolved by resumming some self-energy contributions to the Goldstone propagator amounting to a momentum-independent shift of the Goldstone mass\footnote{For later developments and applications of this resummation, see \cite{2PI,KP,BG}.  Besides resolving the IR issues
just mentioned, it has been shown \cite{EGK} that this resummation also fixes a problem with residual gauge dependence in radiatively generated vacua \cite{Sch}.}
\be
G\rightarrow \Gb = G +\Delta\ .
\label{Gbar}
\ee
One way to determine the right mass-shift $\Delta$ to be used in the resummation is constructive and proceeds by calculating order by order in the perturbative expansion of the potential (or the minimization equation $V'=0$) what  $\Delta$  should be used to remove all infrared problematic terms. One obtains an explicit perturbative result
\be 
\Delta= \kappa \Delta_1+ \kappa^2 \Delta_2+...
\label{Deltapert}
\ee
The fact that this procedure works at all is non-trivial, since for instance the one-loop self-energy term $\kappa \Delta_1$ used has to cancel different IR divergences in the potential at all orders starting at two loops. This procedure was the one used in \cite{MartinIR,V3}. 

Although initially based on the same approach outlined above, \cite{USIR} argued that there is a definite prescription to calculate the  needed  $\Delta$  by integrating out the heavy degrees of freedom, in the spirit of an effective field theory approach. The self-energy diagrams that give $\Delta$ involve only heavy fields  (with masses that do not vanish as $G\rightarrow 0$) or Goldstones (plus photons and gluons) with large momentum. We denote these contributions to the Goldstone self-energy as the hard part, and the method of regions \cite{MoR,MoReview} can be used to make this definition precise.   In principle, these two approaches lead to the same $\Delta$ up to terms that are subleading in powers of $G/X$.  The aim of the present paper is to prove to all orders in perturbation theory that the potential IR issues are removed when $\Delta$, the zero-momentum hard-part of the Goldstone boson self-energy,  is resummed. 

The proof is presented in Sec.~\ref{sec:proof} and an explicit check in the SM at three-loop order is performed in Sec.~\ref{sec:3L}. The discussion of our results can be found in Sec.~\ref{sec:sum}.
Appendices \ref{sec:cubic}-\ref{sec:regions} review known results 
that we include to present a self-contained discussion. App.~\ref{sec:cubic} deals with the cubic coupling of the Goldstone bosons, App.~\ref{sec:over} covers combinatoric issues in the two-particle irreducible (2PI) effective action and App.~\ref{sec:regions} briefly describes the method of regions. The remaining Appendices contain
detailed results for the hard and soft splitting of two-loop contributions 
to the effective potential using the method of regions. This splitting is needed to calculate $\Delta$ as will be explained in Sec.~\ref{sec:proof} and illustrated in Sec.~\ref{sec:3L}. For this task, an expansion in powers of $G/X$ and $(d-4)/2=\epsilon$ is enough but we go beyond this and also perform the splitting of two-loop vacuum integrals for general $d$ and without expanding in $G$.

\section{Proof of Resummation to All Orders \label{sec:proof}} 

In this section, we provide a general proof of how to remove infrared  divergences from the effective potential by means of resummation. The statement we want to prove is the following:

\vskip 0.2 cm

{\em IR problematic terms in the effective potential can be resummed 
by a shift of the Goldstone mass, $G\rightarrow \Gb = G +\Delta$. The shift $\Delta$ is the zero-momentum limit of Goldstone self-energy diagrams that contain only heavy particles and the hard momentum region of light degrees of freedom. The split into soft and hard momenta is made precise by the method of regions.}

\vskip 0.2 cm

Consider the whole set of vacuum 1-particle-irreducible (1PI) diagrams that contribute to the effective potential in the usual perturbative expansion and focus on those that contain Goldstone lines. Following \cite{USIR}, each Goldstone line/propagator can be split as being a $G_s$ (a soft Goldstone, carrying momentum $p^2\sim G$) or a $G_h$ (a hard Goldstone, carrying momentum $p^2\sim X\gg G$, where $X$ represents some nonzero squared mass). We are after those Goldstone contributions that cause IR problems so we are interested in diagrams containg $G_s$ lines and can consider $G_h$ lines on the same footing as propagators of heavy fields.

\begin{figure}[t]
\begin{center}
\includegraphics[width=\textwidth]{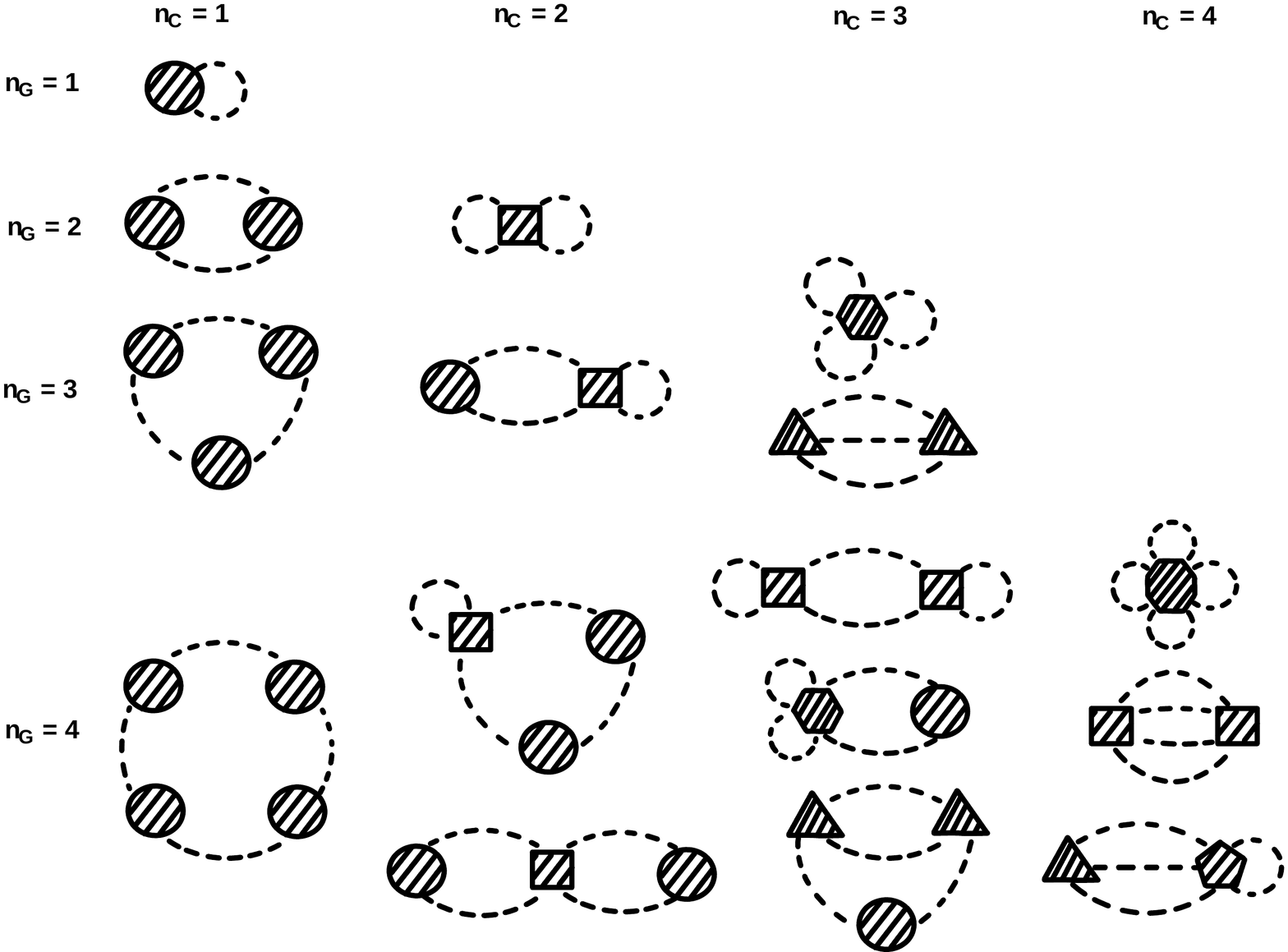}
\end{center}
\caption{\label{fig:DiaGs} \em 
Vacuum diagrams classified according to the number of soft Goldstone (dashed) lines $n_G$ (up to $n_G=4$) and number of Goldstone cycles $n_C$. Blobs represent any subdiagram not involving soft Goldstones, photons or gluons. Different shapes of these blobs are used to distinguish the number of Goldstone lines they have attached.
}
\end{figure}

\begin{figure}[t]
\begin{center}
\includegraphics[width=\textwidth]{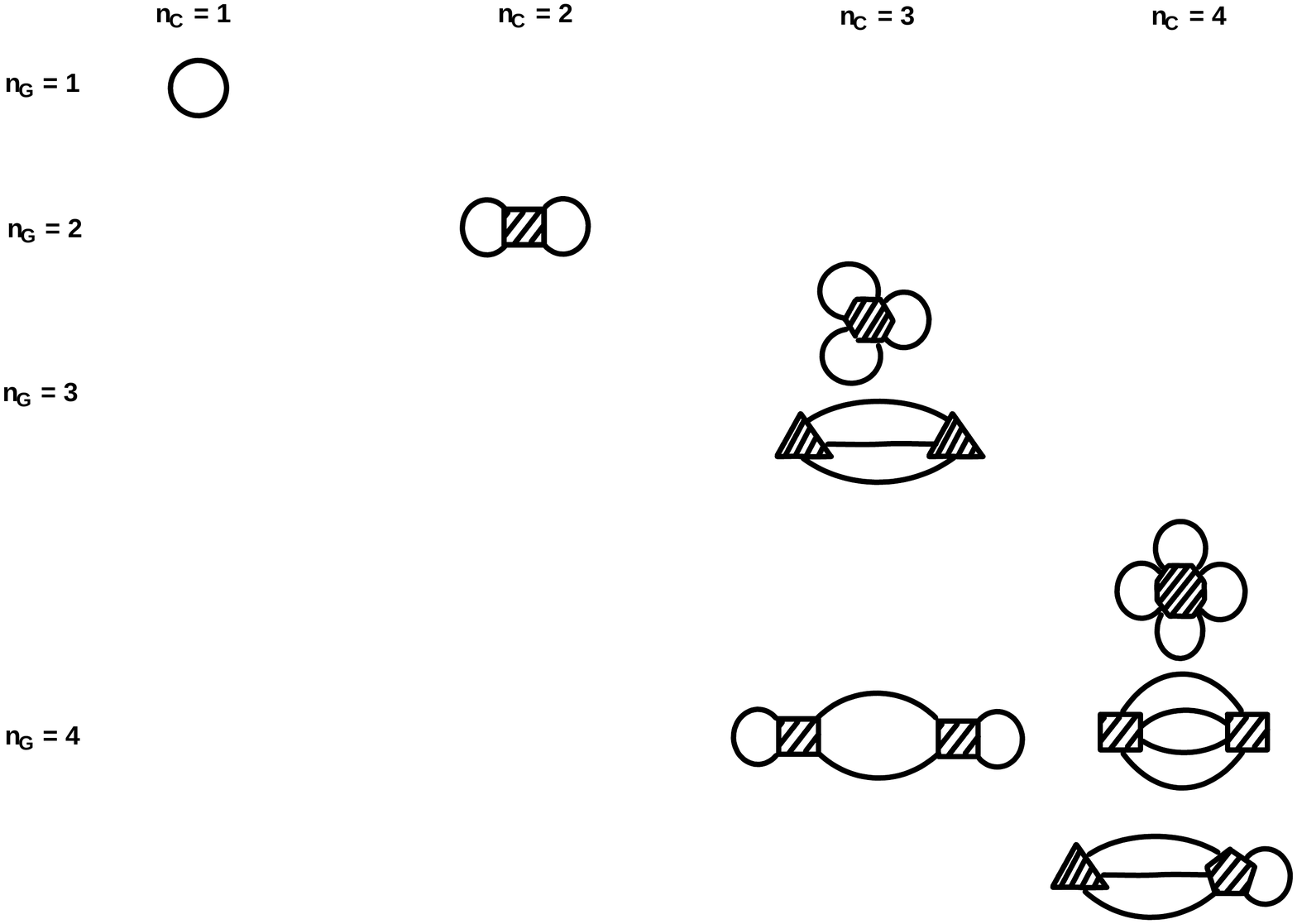}
\end{center}
\caption{\label{fig:DiaGsR} \em 
Vacuum diagrams after resummation of soft Goldstone propagators,
indicated by continuous lines. As in Fig.~\ref{fig:DiaGs}, blobs represent any subdiagram not involving soft Goldstones, photons or gluons.  
}
\end{figure}

After the previous soft/hard splitting one can classify any diagram by the number $n_G$ of $G_s$ lines it carries,
see Figure~\ref{fig:DiaGs}. In this figure we use a dashed line for soft Goldstone propagators, without distinguishing between the three of them ($\chi_i$, $i=1,2,3$). We also classify diagrams according to the number $n_C$ of soft Goldstone cycles they contain. Two $G_s$ lines are defined to be on the same cycle if they necessarily carry the same momentum. Single $G_s$ lines not falling in the previous category 
are considered as cycles by themselves.

At $n_G=0$ we simply have no Goldstone IR divergences and we do not show any diagram.
At $n_G=1$ there is one single topology, with a $G_s$ line as a handle attached to a blob that represents any tangle of lines with hard or heavy propagators only.\footnote{There are no other topologies for $n_G=1$: for instance, a single $G_s$ line between two blobs is not 1PI.} 
Using the method of regions we know that attaching a
$G_s$ handle to a blob pays the price of a $G L_G$ factor, so
these diagrams contribute a term to the potential that scales as $G L_G$. For notational simplicity we use the shorthand notation $L_G\equiv \log(G/Q^2)$, with $Q$ the $\overline{\rm MS}$ renormalization scale.
To see how this comes about, let us write the contribution of the
$n_G=n_C=1$ diagram as
\be
\propto \int_p \frac{1}{G-p^2}\Pi(p)\ ,
\label{nG1nC1}
\ee
where $\Pi(p)$ represents the contribution from the blob and
\be
\int_p\equiv \mu^{2\epsilon}\int \frac{d^dp}{(2\pi)^d i}\ ,
\label{Intp}
\ee
with $d=4-2\epsilon$ and $Q^2=4\pi e^{-\gamma_E}\mu^2$. 
For a soft Goldstone, the method of regions instructs us to leave the Goldstone propagators unexpanded but expand $\Pi(p)$ in powers
of $p^2/X$. As the blob contains only heavy particles or heavy momentum lines this soft-momentum expansion takes the form 
$\Pi(p) = \Pi(0) + {\cal O}(p^2/X)$, and the integral (\ref{nG1nC1}) gives (after renormalization)
\be
\kappa\, G(L_G-1)\,\Pi(0) +{\cal O}(G^2/X)\ ,
\ee
confirming the $GL_G$ scaling mentioned above.

Diagrams with  $n_C=1$ and increasing values of $n_G$ simply correspond to the addition of heavy blobs in the $G_s$ line of 
the $n_G=n_C=1$ diagram. This full series of $n_C=1$ diagrams is taken care of by the usual resummation of the mass in a Goldstone ring and are resummed into a term $\sim\Gb{}^2 L_{\Gb}$ \cite{MartinIR,USIR}. From our re-organization of the perturbative expansion it is also clear that the Goldstone shift in this resummation is precisely of the form (\ref{Gbar}) with $\Delta$ given by hard contributions only and 
defined at zero momentum. The resummed diagram, with the corrected Goldstone propagator (soft and with mass $\Gb$) is represented by a continuous line and shown as the $n_G=n_C=1$ diagram in Fig.~\ref{fig:DiaGsR}.

At $n_G= n_C=2$ we have a blob with two $G_s$ handles, and this gives $G^2 L_G^2$ contributions to the potential, which cause no divergence in $V$ or $V'$, so they are IR safe.\footnote{This result concerning the IR divergence holds irrespective of whether the two Goldstone lines are attached in a planar diagram (as shown)  or in a non-planar way (diagram not shown, for simplicity).} Again, inserting blobs in the $G_s$ lines gives higher $n_G$ diagrams without changing $n_C$ and all of them are resummed into the $n_G=n_C=2$ diagram of Fig.~\ref{fig:DiaGsR}, with exactly the same shift $\Delta$ of the Goldstone mass, so that the resummed diagram, see Fig.~\ref{fig:DiaGsR}, scales as $\Gb{}^2 L_{\Gb}^2$.

The case $n_C=3$ is more interesting as there is more than one topology to deal with. As in previous cases, series of higher $n_G$
diagrams can be resummed and one ends up with the three resummed diagrams shown in Fig.~\ref{fig:DiaGsR}. One can have three $G_s$ loops and this diagram scales as $(\Gb L_{\Gb})^3$.  As the potential has dimension of mass to the fourth power, the $\Gb{}^3$ factor is compensated by the negative dimension of the vertex, which has to scale as $1/X$ (as it should, being a six-legged vertex). 

At $n_C=3$ we also have  the diagram with two cubic blobs exchanging 3 $G_s$ lines. To see how this scales with $\Gb$, we can
estimate from the method of regions that the two-loop integrals with
three soft Goldstones can give a factor $\sim \Gb L_{\Gb}^2$. If the cubic vertex were of order $\sqrt{X}$ we would end up with a contribution to the potential of order $X \Gb L_{\Gb}^2$ that would cause IR divergences.
Clearly, the scaling of the $n$-legged $G_s$ vertex is relevant 
for the success of the resummation program, and we discuss it next.

We are interested in particular in the small-momentum expansion of the $G_s$ vertices. For  our reorganized perturbative expansion, by construction, only heavy species and hard light ones (Goldstones, photons and gluons) contribute to these vertices.
The zero-momentum values of the $G_s$ vertices can be obtained from derivatives of the effective potential by exploiting the constraints that gauge invariance imposes on such field derivatives \cite{WeinbergVder}. Let us write the Higgs doublet as
\be
H = \frac{1}{\sqrt{2}}\left(
\begin{array}{c}
\chi_1+ i \chi_2\\
h+\phi +i \chi_3
\end{array}
\right)\ ,
\ee
where the $\chi_i$ fields are the three Goldstones, taken as real fields (alternatively we have $\chi^0=\chi_3$ and $\chi^\pm=(\chi_1\pm i\chi_2)/\sqrt{2}$).
The effective potential in Landau gauge has a global $SU(2)$ symmetry so that it can only be a function of the invariant $|H|^2$
(even after including radiative corrections).
The simplest way to deal with the constraints imposed by gauge invariance on the different scalar interactions is then to consider the SM effective potential as a function of 
\be
|H|^2=\frac12 (h+\phi)^2+\frac12 \chi^2\ , \quad {\rm with}\quad 
\chi^2 = \sum_{i=1}^3\chi_i^2\ .
\ee
The (zero-momentum) Goldstone interactions can then be obtained by expanding the potential in powers of $\chi^2$ around the background value $\phi$. We can write
\be
\label{eq:L_exp}
-{\cal L}_{\chi,p^2=0} \equiv \sum_{n=1}^{\infty}\frac{1}{n!\,2^n}\lambda_{\chi,2n}\chi^{2n} = \sum_{n=1}^{\infty}\frac{1}{n!}\left[\frac{\partial^nV(|H|^2)}{(\partial \chi^2)^n}\right]_{h=0,\chi=0}
\chi^{2n}\ .
\ee
Noting further that $\partial V/\partial \chi^2=(\partial V/\partial\phi)/(2\phi)$, we arrive at
\be
\lambda_{\chi,2n} = \left(\frac{1}{\phi}\frac{\partial}{\partial\phi}\right)^n V(\phi)\ .
\label{Gvertices}
\ee
Notice that this implies that vertices with an odd number of Goldstone legs, and the cubic vertex in particular, vanish. We provide an alternative and more general proof that the cubic vertex vanishes in Appendix~\ref{sec:cubic}.

As a cross-check of (\ref{Gvertices}), for the tree-level potential in Eq.~(\ref{V0}) one gets
\bea
G= \lambda^{(0)}_{\chi,2}&=& \frac{1}{\phi}\frac{\partial V_0}{\partial\phi}=-m^2+\lambda \phi^2\ ,\nonumber\\
 \lambda^{(0)}_{\chi,4}&=& \frac{1}{\phi}\frac{\partial \lambda_{\chi,2}}{\partial\phi}=2\lambda\ , \nonumber\\
  \lambda^{(0)}_{\chi,2n}&=& 0 \quad ({\rm for}\, n\geq 3)\ .
\eea
The radiative corrections to these tree-level results follow directly 
from using the radiatively corrected effective potential, and all
$\lambda_{\chi,2n}$ become nonzero, although it continues to be true that vertices with an odd number of Goldstone legs vanish.

Before proceeding with the calculation of soft-Goldstone vertices, notice that the blobs in Figs.~\ref{fig:DiaGs} and \ref{fig:DiaGsR} can contain additional pieces besides those obtained from the effective potential via the relation (\ref{Gvertices}). The reason is that (\ref{Gvertices}) only contains contributions that are 1PI, while
the blobs that appear from heavy particles can also contain one-particle-reducible (1PR) contributions. As a simple example, 
consider a contribution to the vertex in the $n_G=n_C=2$ diagram
of Fig.~\ref{fig:DiaGsR} from a $T$-channel exchange of a Higgs.
However, our proof is only based on the mass scaling of Goldstone vertices, which is not changed by these contributions. Moreover, 
such 1PR contributions cannot be present at zero-momentum
for the quadratic and cubic Goldstone couplings as they would require the exchange of a heavy or hard particle with the same quantum numbers as the Goldstones. The only candidates available are
derivatively coupled longitudinal gauge bosons, but the derivative introduces a momentum dependence that goes to zero with the external momentum. For odd Goldstone vertices, also 1PR contributions
can only involve exchanges of derivatively coupled gauge bosons
as the exchange of a Higgs would require a coupling between an odd number of Goldstones and a Higgs and this also vanishes at zero momentum, as is obvious from the previous discussion.

In our particular setting, as only hard propagators contribute now to the vertices, one needs to use the radiatively corrected potential with only hard particles in the loops when using (\ref{Gvertices}).  We will call such potential $V_{\rm hard}$.
Starting from the lowest dimension $G_s$ vertex, we get\footnote{In our procedure, the fact that all Goldstone bosons $\chi_i$ receive the same mass shift (so that all of them are massless at the true vacuum) is built in from the start. Putting back indices we would write $\Delta_{ij}=\Delta \delta_{ij}$. }
\be
\Gb = \lambda_{\chi,2}= \frac{1}{\phi}\frac{\partial V_{\rm hard}}{\partial\phi}=G+\Delta\ ,
\label{Delta}
\ee
which gives a concrete calculational definition of the shift $\Delta$. Generically, we expect that $\Delta$ scales as $X$. As $V_{\rm hard}$ also contains hard-Goldstones, $G$ can also appear in $\Delta$.  The method of regions instructs us to expand the propagators of hard Goldstones in powers of $G/p^2$ and so, $G$ can only appear in the mass scaling of these vertices with positive powers:
\be
\Delta\sim X +{\cal O}(G)\ .
\label{O2}
\ee
Such $G$ dependence is absent in $\Delta_1$ and the one in $\Delta_2$ is only relevant to resum IR divergences at four-loop order.

For the cubic vertex (or any odd vertex) we obtain zero at vanishing external  momentum, so that the only possible contributions to this vertex will be proportional to the external momentum.\footnote{Notice that out of $H$ and $D_\mu H$ one can now build operators that give rise to such couplings, {\it e.g.} $c\, \partial^\mu |H|^2(H^\dagger D_\mu H) + {\rm H.c.}$, where $c$ is a complex constant (with mass dimension $-2$).} As the external legs are soft Goldstones, this provides only additional positive powers of $G$ and we conclude that  
\be
\lambda_{\chi,3}\sim \sqrt{G}\ .
\label{O3}
\ee 
Therefore, the $n_G=n_C=3$ resummed diagram of Fig.~\ref{fig:DiaGsR} with two cubic vertices also scales as $\Gb{}^2L_{\Gb}^2$ and is IR safe. 

Let us consider next the quartic coupling $\lambda_{\chi,4}$. This is dimensionless and radiative corrections to it can only depend on other dimensionless couplings, ratios $X/Y$ of heavy squared masses
or positive powers of $G/X$. The $n_G=n_C=3$ diagram with two such quartic couplings in Fig.~\ref{fig:DiaGsR} therefore scales like $\Gb{}^2L_{\Gb}^3$ and is also IR safe.

After these concrete examples of resummed diagrams we are ready for the generalization to arbitrary topologies. Let us call $\gamma(P,V,L)$ an arbitrary resummed diagram with a number $P$ of Goldstone propagators, a total number $V=\sum_n (V_{2n+2}+V_{2n+1})$ of vertices (with $V_{2n+i}$ the number of vertices with $2n+i$ legs, with $n\geq 1$) and a number $L$ of loops.
These numbers
are related by the identity
\be
P-V=L-1\ .
\label{PVL}
\ee
Let us calculate how such a generic diagram scales with $\Gb$. Each loop integral brings a power $\Gb{}^2$, each Goldstone propagator a power $1/\Gb$. The Goldstone vertices scale
as
\bea
\lambda_{\chi,2n+2}&\sim &X^{1-n}\left[1+{\cal O}(\Gb/X)\right]\ ,\nonumber\\
\lambda_{\chi,2n+1}&\sim &\Gb^{1/2} X^{1-n}\left[1+{\cal O}(\Gb/X)\right]\ .
\label{On}
\eea
We therefore find the scaling\footnote{The dependence on $L_{\Gb}$ always comes from integration over the loops of soft Goldstones as hard Golstones inside 
vertices give analytic contributions. In (\ref{scaling}) we write $L_{\Gb}$ raised to the highest possible power from an $L$-loop diagram but it should be understood that lower powers also appear.
}
\bea
\gamma(P,V,L)&\sim &\Gb{}^{2L-P+\sum_n V_{2n+1}/2}X^{V-\sum_n n (V_{2n+2}+V_{2n+1}) }(L_{\Gb})^L\nonumber\\
&=& \Gb{}^{2L-P}X^{2V-P}(L_{\Gb})^L\left(\frac{\Gb}{X}\right)^{\sum_nV_{2n+1}/2}\ ,
\label{scaling0}
\eea
where the last equality follows from the relation $P=\sum_n \left[(n+1) V_{2n+2}+(n+1/2)V_{2n+1}\right]$. Notice that, using (\ref{PVL}), it can be checked that 
(\ref{scaling0}) has the right [mass]$^4$ dimension. In fact, using (\ref{PVL}) we can eliminate $L$ and write the simple expression:
\be
\gamma(P,V,L)\sim \Gb{}^{2}(L_{\Gb})^L \left(\frac{\Gb}{X}\right)^{P-2V+\sum_nV_{2n+1}/2}\ .
\label{scaling}
\ee

As the vertices 
involved in the resummed diagrams always have more than 2 legs ($n\geq 1$), one can derive the inequality
\be
P-2V+\frac12 \sum_nV_{2n+1}=\sum_n (n-1) \left(V_{2n+2} + V_{2n+1}\right)\geq  0 \ ,
\ee
and this implies that (\ref{scaling}) is IR safe for $\Gb\rightarrow 0$.

It is transparent that the following result follows directly from the scaling of the Goldstone couplings, which is dictated by symmetry and dimensional analysis as is customary in effective theory (EFT) approaches, in this case an effective theory  for soft Goldstones. The renormalizable two point function is sensitive to heavy scales and one gets the scaling (\ref{O2}). A global $SU(2)$ symmetry protects the cubic coupling, that scales as in (\ref{O3}) and higher order operators are suppressed by the heavy scale $X$ as given in (\ref{On}).    The only physical mass in the EFT is the mass of the Goldstone, $\Gb$.  
 Any IR problem had to result from interaction vertices that come with a coupling with $X$ raised to a positive mass dimension, but this, which was only possible for the cubic coupling does not happen.
In the previous discussion we have ignored the presence of photons and gluons, that should also be taken into account in the effective theory and also contribute to 
the soft potential besides the diagrams in Figs.~\ref{fig:DiaGs}
and \ref{fig:DiaGsR}. However, dimensional analysis shows again
that these contributions cannot generate Goldstone IR divergences either.

Let us close this section with some comments on the proof we have given, before we present the explicit three-loop check of our approach in the following section.
First, many resummation schemes suffer from over-counting problems. This is most transparent in the 2PI effective action that contains one additional term that compensates for contributions that are over-counted in the resummation. Our prescription does not suffer from over-counting, essentially because we do not resum the contributions from soft Goldstone bosons in the self-energy. Details on the over-counting problem are given in Appendix~\ref{sec:over}.

Second, the diagrams in Figs.~\ref{fig:DiaGs} and \ref{fig:DiaGsR} might suggest that we resum the full momentum-dependent hard part of the self-energy. While the leading term (the mass renormalization) and the next-to-leading term (the wavefunction renormalization) in a momentum expansion can be easily resummed, resumming the full expression would be technically very demanding. However, notice that re-expanding the propagators in any subleading terms will not lead to IR issues and hence resumming the leading $p^2=0$ contribution is in fact sufficient to resolve the IR problems.

\section{Cross Check of Resummation at Three Loops \label{sec:3L}} 

In this section we check that our prescription for resumming soft-Goldstone contributions works to make safe the SM potential at the three-loop level (as calculated in \cite{V3}). We also compare our 
result with the resummation procedure performed at the same loop level
in \cite{V3}. As explained in the previous sections, one key ingredient
for the resummation is the definition of the shift of the Goldstone mass,
$G\rightarrow \Gb = G +\Delta$, with $\Delta$ defined by Eq.~(\ref{Delta}). 

The perturbative expansion of $\Delta$ starts at one loop, as in (\ref{Deltapert}). The hard part of the one-loop potential, $V_{1,{\rm hard}}$, includes contributions from all massive particles except the Goldstone bosons (that is, the contribution of hard Goldstones vanishes), while the soft part, $V_{1,{\rm soft}}$, just comes from Goldstone loops.
The contribution of each massive species to the one-loop potential is
\be
\delta_\alpha V_1 = \frac14 N_\alpha X_\alpha^2\left(L_{X_\alpha}-C_\alpha\right)\ ,
\ee
where for each particle species $\alpha$,  $X_\alpha=M_\alpha^2(\phi)$ is the field-dependent squared mass, $N_\alpha$ counts the number of degrees of freedom (taken negative for fermions) and $C_\alpha$ is a constant equal to 3/2 for scalars or fermions and equal to  5/6 for gauge bosons.
One then obtains \cite{MartinIR,USIR}
\be
\Delta_1 = \frac{1}{\phi}\frac{\partial V_{1,{\rm hard}}}{\partial\phi}= -6 h_t^2 A(T) +
3\lambda A(H) 
 + \frac12 g^2 [3A(W) +2W]
+ \frac12  g_Z^2 [3A(Z) +2Z] \, ,
\label{Delta1}
\ee
where we use $g_Z^2=g^2+g'{}^2$ and
\be
T=\frac12 h_t^2\phi^2\ ,\quad H=-m^2+3\lambda\phi^2\ , \quad 
W=\frac14 g^2 \phi^2\ , \quad Z=\frac14 g_Z^2 \phi^2\ ,
\ee
are the (squared) masses of the top quark, the Higgs boson, and the
$W$ and $Z$ gauge bosons, respectively, that is $X_\alpha=\{T,H,W,Z\}$. The (renormalized) one-loop function $A$ is defined
as
\be
A(X)=X (L_X-1)\ .
\ee

At two loops, finding $\Delta_2$ requires obtaining the hard piece of
the two-loop effective potential. This task is readily performed by using the method of regions to split the Goldstone bosons into soft and hard ones. Looking at the explicit expression of the SM two loop potential 
as given in \cite{V2,V2M} we see that the (nontrivial) two-loop functions containing
Goldstone contributions are of the following types:
\bea
I(G,0,0)\ , \quad I(G,G,0)\ , \quad
I(G,G,X)\ ,\quad I(G,X,0)\ , \quad I(G,X,X)\ ,\quad  I(G,X,Y)\ ,
\label{Is}
\eea
where $X,Y$ represent any other massive particle and $I(m_1^2,m_2^2,m_3^2)$ is the (renormalized) two-loop function
to which all two-loop functions can be reduced. It
corresponds to the setting-sun diagram with three scalar propagators
with the indicated masses and a precise definition can be found in the Appendix E. Here we follow the notation of \cite{V2M}, where explicit
expressions for $I$ can also be found. 

By using the method of regions, as detailed in Appendices F-H,
one can split the integrals in (\ref{Is}) as
\bea
I(G,0,0)&=&I(G_s,0,0)+I(G_h,0,0)\ ,\nonumber\\
I(G,G,0)&=& I(G_s,G_s,0)+2I(G_s,G_h,0)+I(G_h,G_h,0)\ , \nonumber\\
I(G,G,X)&=&I(G_s,G_s,X)+2I(G_s,G_h,X)+I(G_h,G_h,X)\ , \nonumber\\
I(G,X,0)&=&I(G_s,X,0)+I(G_h,X,0)\ , \nonumber\\ 
I(G,X,X)&=&I(G_s,X,X)+I(G_h,X,X)\ \ , \nonumber\\  
I(G,X,Y)&=&I(G_s,X,Y)+I(G_h,X,Y)\ ,
\eea
where $G_s$ ($G_h$) denote a soft (hard) Goldstone. Explicit
expressions for these split integrals are given in Appendices F-H,
both for general dimension $d=4-2\epsilon$ and in an expansion
in powers of $G$ and $1/\epsilon$. The last form is all that is needed for our purposes.

The Goldstone contribution to the hard part of the two-loop potential $V_{2,{\rm hard}}$ is given by the $I$ functions involving only $G_h$, as above, while the soft part, $V_{2,{\rm soft}}$, comes from contributions containing $G_s$. 
The two-loop shift of the  Goldstone mass is then calculated as
\be
\Delta_2 = 
\frac{1}{\phi}\frac{\partial V_{2,{\rm hard}}}{\partial\phi}\ .
\label{Delta2}\ee

To check that the shift $\Delta$ defined in this way resums the IR
divergent pieces of the three-loop potential it is enough to calculate
$\Delta$ up to two-loop order and order ${\cal O}(G^0)$. The result of our calculation for $\Delta_2$  is given in Appendix~I. In order to show explicitly that this 
indeed works, we proceed as follows. We use the full three-loop Standard Model effective potential of~\cite{V3}. The soft parts of the effective potential, which are those that contain the IR divergent terms, take the following form. At one-loop, one has
\be
V_{1,{\rm soft}} = G^2(a_1 L_G +b_1)\ ,
\ee
with 
\be
a_1=\frac34\ ,\quad
b_1=-\frac98\ .
\ee
At two loops, we find that, in an expansion in powers of $G/X$, the soft part of the potential can be written as:
\be
V_{2,{\rm soft}} =2 \Delta_1 G \left[a_1(L_G+1/2)+b_1\right] + G^2(a_2 L_G^2 +b_2 L_G+c_2)
+{\cal O}(G^3/X)\ ,
\ee
with 
\bea
a_2&=&\frac98 (2\lambda-e^2)\ ,\\
b_2&=&\frac38\left\{4 h_t^2 (3L_T-1)-2\left(g_Z^2-3e^2\right)L_Z-4g^2L_W+\frac{8g^4}{g'{}^2}(L_Z-L_W)+8g^2-\frac{11 g^4}{g_Z^2}
\right.\nonumber\\
&&+\frac{5g_Z^2}{2}
-16\lambda-2g^2\frac{\left(g^2 L_W-8\lambda L_H\right)}{(g^2-8\lambda)}-\left. g_Z^2\frac{\left(g_Z^2 L_Z-8\lambda L_H\right)}{(g_Z^2-8\lambda)}\right\}\ ,
\eea
where $e=g g'/g_Z$.
Finally, at three-loop order we find that the soft potential can be expressed as
\bea
V_{3,{\rm soft}}&=& 2G\left[a_2 \Delta_1 (L_G^2+L_G)
+(b_2 \Delta_1 +a_1 \Delta_2)(L_G+1/2)+c_2 \Delta_1 +b_1 
\Delta_2 \right]
\nonumber\\
&+&
\Delta^2_1 \left[a_1 (L_G+3/2)+b_1\right]
+{\cal O}(G^2)\ .
\eea

From the formulas above it is obvious that the soft potential 
\be
V_{\rm soft}=\kappa V_{1,{\rm soft}}+\kappa^2 V_{2,{\rm soft}}+
\kappa^3 V_{3,{\rm soft}}+...
\ee
has IR divergences (and $V'_{\rm soft}$ as well). It is straightforward
to check that the resummed potential
\be
V_{R,{\rm soft}} \equiv \kappa \Gb{}^2(a_1 L_{\Gb} +b_1)+
\kappa^2  \Gb{}^2\left[a_2 (L_{\Gb})^2 +b_2 L_{\Gb}+c_2\right]
\ee
with $\Gb=G+\Delta$,
reproduces the unresummed one upon expansion in powers of $\Delta/G$ but is IR safe.

Finally, we can also compare our two-loop result for $\Delta$ in (\ref{Delta}) with the shift $\hat\Delta$ used
in \cite{MartinIR,V3} for resummation and obtained using a different procedure (basically examining order by order the IR divergences of the minimization condition for the Higgs potential to identify the shift required to resum them).  To make contact with the result for $\hat\Delta$ as given in \cite{MartinIR,V3} we need to expand the Higgs contribution to our $\Delta_1$, given in (\ref{Delta1}), in powers of $G$ (using $H=2\lambda\phi^2+G$)
\be
\Delta_1(H) = \Delta_1(2\lambda\phi^2) + G\, \partial \Delta_1/\partial H + {\cal O}(G^2)\ ,
\ee
and then perform the substitution $G\rightarrow G+\Delta$. Identifying
order by order the different contributions to $\Delta$ and $\hat\Delta$
we find that agreement up to two loops requires the relations
\bea
\hat\Delta_1 &=& \Delta_1(2\lambda\phi^2)\ ,\nonumber\\
\hat \Delta_2 &=&
\Delta_2 + \Delta_1 \partial \Delta_1/\partial H \ ,
\eea
which are indeed satisfied.

\section{Summary and Discussion\label{sec:sum}} 

We have shown that the IR divergences of the Landau gauge effective potential  from Goldstone bosons can be removed by resumming zero-momentum self-energy diagrams of soft Goldstone boson lines. The appropriate self-energy is given by the hard contributions, meaning either contributions from heavy particles or contributions from 
light particles with  hard momentum. The split into hard and soft modes is implemented and made precise by the method of regions. We provided a diagrammatic proof to all orders and also an explicit check using the effective potential in the SM at three-loop order.

As a guideline, we follow the principles of an effective action in the Wilsonian sense, where only the soft momentum 
contributions of the light particles are relevant at low energies, completing the analysis started in \cite{USIR}. 
Our proof provides an explicit prescription for which contributions to the Goldstone boson self-energy have to be resummed to resolve the IR issues. This puts in firmer ground the procedure used in earlier work \cite{MartinIR,V3} to perform this resummation. For example, in these papers it is argued that terms of order  $G \, L_G$ should not be resummed which agrees with our findings. However, there are terms for which it is a priori unclear if they have to be resummed or not. For instance, there are terms of order $G$ in the hard part as well as the soft part of the Goldstone self-energy and according to our arguments, which provide a rationale to distinguish these terms, only the former have to be resummed.
Ultimately, the reason why this approach to resummation works and clarifies the picture is due to symmetry and dimensional analysis, the usual ingredients for the `magic' of effective theory approaches.

Finally, let us mention that the Goldstone boson catastrophe is not the only application of the 
methods presented here. Most theories with several scales and light scalars suffer from the hierarchy problem.  
This means that all scalars have self-energy contributions of order of the heavy scale and they have to be absorbed by the tree level mass to obtain a light scalar.
In the $\overline{\mathrm{MS}}$ scheme, the tree level mass of the light boson and its self-energy are then of the same order
and techniques along the lines presented in this work have to be used to save or improve the convergence of perturbation theory.

\section*{Acknowledgments}

J.R.E. thanks LPTHE, Paris for hospitality and Johannes Braathen and Mark D. Goodsell for discussions that triggered this work and for participating in its early stages. T.K. acknowledges support by the German Science
Foundation  (DFG)  within  the  Collaborative  Research  Center  (SFB)  676  Particles,  Strings
and the Early Universe.
The work of J.R.E. has been partly supported by the ERC
grant 669668 -- NEO-NAT -- ERC-AdG-2014, the Spanish Ministry MINECO under grants  2016-78022-P and
FPA2014-55613-P, the Severo Ochoa excellence program of MINECO (grant SEV-2016-0588) and by the Generalitat grant 2014-SGR-1450.

\appendix

\section{The Cubic Goldstone Coupling \label{sec:cubic}} 

In this Appendix we give a more general proof for the vanishing of
the cubic Goldstone coupling at zero external momentum, following~\cite{WeinbergVder,BG}.

We consider the effective potential in Landau gauge and denote all scalar fields collectively as $\phi_i$. When the electroweak symmetry is spontaneously broken, a global symmetry arises for the Goldstone bosons. That means that there are $M$ transformations $\phi_i \to \bar \phi_i = \phi_i + \epsilon_i^m$ ($m \in [1,M]$)  that leave the effective potential invariant, $V(\phi_i) = V(\bar \phi_i)$. 

This implies
\be
\epsilon_i^m\frac{\partial V}{\partial\phi_i} = 0 \, ,
\ee
and also the first derivative of this relation 
\be
\frac{\partial\epsilon_i^m}{\partial\phi_j} \frac{\partial V}{\partial\phi_i} +
\epsilon_i^m\frac{\partial^2 V}{\partial\phi_i \partial\phi_j} = 0 \, .
\ee
The interpretation of the last equation is that in the vacuum (where $\partial V/\partial \phi_i = 0$) there are $M$ linear combinations in scalar fields 
that are massless Goldstone bosons, $\chi_m$ with $\langle\chi_m|\phi_i\rangle=\epsilon_i^m$. The subsequent derivative of this equation reads 
\be
\frac{\partial\epsilon_i^m}{\partial\phi_j \partial\phi_k} \frac{\partial V}{\partial\phi_i} +
\frac{\partial\epsilon_i^m}{\partial\phi_k} \frac{\partial^2 V}{\partial\phi_i \partial\phi_j} +
\frac{\partial\epsilon_i^m}{\partial\phi_j} \frac{\partial^2 V}{\partial\phi_i \partial\phi_k} +
\epsilon_i^m\frac{\partial^3 V}{\partial\phi_i \partial\phi_j \partial\phi_k} = 0 \, .
\ee
Contracting this relation with the vectors $\epsilon_i^m$ one finds
\be
\epsilon_i^m\epsilon_j^n\epsilon_k^o\frac{\partial^3 V}{\partial\phi_i \partial\phi_j \partial\phi_k} = 0 \quad \textrm{(in vacuum) .}
\ee
In conclusion, the cubic Goldstone couplings are of the order $p^2/X$ and $G/X$ where $X$
is some hard mass scale and $p$ denotes the scale of external momenta. Hence, the cubic coupling
cannot give rise to IR issues in the EFT.

\section{The Over-Counting Problem \label{sec:over}} 

In this Appendix we discuss the combinatorics behind the resummation in the two-particle-irreducible (2PI) effective action, and in particular the over-counting problem. The original proof was given using functional methods~\cite{Cornwall:1974vz}. Here we will use the more diagrammatic proof used in~\cite{deDominicis:1964zz, Blaizot:2003an}. We use Euclidean signature in this section following~\cite{Blaizot:2003an}.

Imagine that the full self-energy of a field was known. In this case, the self-energy can be added to the propagator and subtracted again as a counter term. Diagrams that are two-particle-reducible (2PR) contain a self-energy subdiagram and belong to a set of diagrams that is ultimately canceled by the additional diagram generated by the counterterm.

\begin{figure}[t]
\begin{center}
\includegraphics[width=0.4\textwidth]{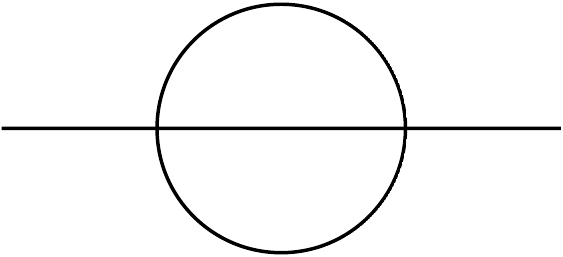}
\end{center}
\caption{\label{fig:phi4} \em 
Closing the external legs of this self-energy diagram leads to an enhanced (four-fold) symmetry. This upsets the naive counting in the 2PI formalism and leads to the over-counting problem.
}
\end{figure}

One might argue that also 2PI diagrams can be obtained by closing the two legs of a self-energy diagram. 
However, if a vacuum diagram is obtained this way, this cancellation is eventually incomplete. This is due to the fact that the generated vacuum diagram can have an enhanced symmetry (see Fig.~\ref{fig:phi4}). This is then compensated by the second term in the 2PI relation
\be
V_{eff} = \frac12 {\rm Tr}\, \log D^{-1} - \frac12 {\rm Tr} \, \Pi\, D   + \Phi \, \label{eq:2PIre} \, .
\ee

The combinatorics behind this resummation can be understood~\cite{deDominicis:1964zz, Blaizot:2003an} by introducing the concept of a cycle.\footnote{This deviates from the definition used in~\cite{Blaizot:2003an} where a cycle contains at least two propagators with common momentum and at least one self-energy insertion.}  A cycle denotes a chain of propagators (with the same momentum) with 2PR self-energy insertions. Notice that by our definition, all propagators are part of a cycle. 
Using the definitions 
\begin{flushleft}
\quad $n_C$ : number of cycles in the diagram \\
\quad $n_S$ : number of skeleton diagrams obtained by removing self-energies \\
\quad $n_L$ : number of lines in cycles
\end{flushleft}
one finds for the logarithmic contribution to the effective potential the 
relation
\be
\label{eq:conA}
\frac12 { \rm Tr} \log \left( 1 + D_0 \, \Pi \right) 
= \sum_\gamma n_C(\gamma) \, w(\gamma) \, , 
\ee
where $w(\gamma)$ denotes the value of the vacuum diagram $\gamma$. Likewise, for the second contribution to the effective action one finds the relation
\be
\label{eq:conB}
- \frac12 {\rm Tr} \,  \Pi \, D 
 =  \sum_\gamma n_L(\gamma) \, w(\gamma) \, . 
\ee
Finally, the 2PI piece is
\be
\Phi = \sum_\gamma n_S(\gamma) \, w(\gamma) \, . 
\ee
The resummation formula is then based on the relation
\be
n_C - n_L + n_S = 1 \, .  \label{eq:comb_re}
\ee
Ultimately, the proof on the resummation of Goldstone bosons is based on only resumming the hard part of the self-energy of the soft Goldstone bosons. Hence, only diagrams with $n_c = 1$ are resummed into the logarithmic Coleman-Weinberg contribution. All contributions to $\frac12 {\rm Tr} \, \Pi\, D$ then cancel against the corresponding contributions in $\Phi$ thanks to $n_L = n_S$. This leaves only terms with $n_C > 1$ in $\Phi$. In other words: since the self-energies that are resummed do no contain any soft Goldstone boson lines, closing the external legs will not enhance the symmetry of the diagram. Hence, no over-counting problem arises in our resummation.
 
\section{Method of Regions \label{sec:regions}} 

To split the Goldstone self-energy contributions into a hard and a soft part, we apply the method of regions \cite{MoR}. Since the Goldstone self-energy can be obtained by taking derivatives of the effective potential as in (\ref{Delta}) we perform the splitting into hard and soft parts directly in the effective potential contributions.

Consider then a vacuum diagram with different mass scales. In general, we will have light particles (including the Goldstone, gluons and photons) and heavy particles (everything else)
\be
G \ll \Lambda \ll X=\{W, Z, H, T \dots\} \, ,
\ee
with $W=g^2\phi^2/4$, $Z=(g^2+g'{}^2)\phi^2/4$, $H=-m^2+3\lambda\phi^2$ and $T=h_t^2\phi^2/2$.
Next, all momentum integrations are split into soft, $p^2 \ll \Lambda$, and hard, $p^2 \gg \Lambda$, regions, leading to 
\be
\int_M I(p^2) \, d^dp =  \int_S I(p^2) \, d^dp + \int_H I(p^2) \, d^dp \, .
\ee
In the soft regions, the integrand $I(p^2)$ can be expanded in momentum over the large masses, $p^2/X$, while in the hard regions the integrals can be expanded in $G/p^2$. 
We denote these expansions as $I^S(p^2)$ and $I^H(p^2)$:
\be
\int_M I(p^2) \, d^dp =  \int_S I^S(p^2) \, d^dp + \int_H I^H(p^2) \, d^dp \, .
\ee
After the expansion, both integration regions can be extended to the full Minkowski space, yielding 
\be
\int_M I(p^2) \, d^dp =  \int_M I^S(p^2) \, d^dp + \int_M I^H(p^2) \, d^dp 
 -\int_H I^S(p^2) \, d^dp - \int_S I^H(p^2) \, d^dp \, .
\ee
At first sight, this formula seems to lead to some over-counting, but the dimensional regularization actually takes care of this. 
The added regions actually allow in all our cases for a double expansion. 
\be
-\int_H I^S(p^2) \, d^dp - \int_S I^H(p^2) \, d^dp
= -\int_H I^{S,H}(p^2) \, d^dp - \int_S I^{H,S}(p^2) \, d^dp \, .
\ee
As long as these two expansions commute, the added regions can be combined into an integral over all space that vanishes in dimensional regularization
\be
-\int_H I^S(p^2) \, d^dp - \int_S I^H(p^2) \, d^dp
= -\int_M I^{S,H}(p^2) \, d^dp  = 0 \, ,
\ee
due to the fact that the double expansion will lead to scaleless monomials in the momentum~$p^2$.

In summary, the method of regions implies 
\be
\int_M I(p^2) \, d^dp =  \int_M I^S(p^2) \, d^dp + \int_M I^H(p^2) \, d^dp \, ,
\ee
and its generalization to multiple momentum integrals.

\section{Splitting of \bma{{\bf B}(X,G;0)}}

Among the basis integrals that appear in calculating the hard/soft split
of the two loop potential, there appears the one-loop integral 
with two propagators of different mass. The splitting of the one-loop integral [with the integration defined as in (\ref{Intp})]
\be
 {\bf B}(X,G;p^2) =\kappa^{-1} \int_q \frac{1}{(G-q^2)[X-(q+p)^2]}\ ,
\ee
at zero external momentum is straightforward and gives
\be
{\bf B}(X,G_s;0)=\frac{{\bf A}(G)}{X-G}\ ,\quad\quad
{\bf B}(X,G_h;0)=-\frac{{\bf A}(X)}{X-G}\ ,
\ee
with
\be
{\bf A}(Z)=\kappa^{-1}\int_p\frac{1}{Z-p^2}=Z \left[\frac{Z}{4\pi\mu^2}\right]^{-\epsilon}\Gamma(-1+\epsilon)=Z\left[-\frac{1}{\epsilon}+L_Z-1+{\cal O}(\epsilon)\right]\ ,
\ee
with $L_Z\equiv \log(Z/Q^2)$ and $Q^2\equiv 4\pi\mu^2 e^{-\gamma_E}$.
Obviously, the sum of the split parts reproduces the full result:
\be
{\bf B}(X,G;0)={\bf B}(X,G_s;0)+{\bf B}(X,G_h;0)\ .
\ee
After subtracting divergences, the renormalized version of this split reads
\be
B(X,G_s;0)=\frac{A(G)}{X-G}\ ,\quad\quad
B(X,G_h;0)=-\frac{A(X)}{X-G}\ ,
\ee
with $A(Z)=Z(L_Z-1)$, where we distinguish renormalized functions by using normal fonts.

\section{Splitting of \bma{{\bf I}(G,G,X)}}

A basis integral that appears repeatedly in two-loop vacuum
contributions is
\be
{\bf I}(X,Y,Z) = \kappa^{-2}\int_p\int_q\frac{1}{(X-q^2)(Y-p^2)[Z-(p+q)^2]}\ ,
\ee
and in this Appendix and the following ones we perform the splitting
of this integral when it contains Goldstone propagators.

The splitting of ${\bf I}(G,G,X)$ into ${\bf I}(G_s,G_s,X)$, ${\bf I}(G_s,G_h,X)$ and ${\bf I}(G_h,G_h,X)$ can be obtained in a direct way by using the method of regions to expand the propagators and truncating the resulting power series at the desired order in $G$. Some of the resulting momentum integrals are straightforward and others can be evaluated with the help of automated tools like FIRE \cite{FIRE}.
Working up to ${\cal O}(G^2)$, one obtains the compact expressions
\bea
{\bf I}(G_h,G_h,X)&=&{\bf I}(0,0,X)\left[1-2(d-3)\frac{G}{X}+2(d-3)(d-5)\frac{G^2}{X^2}\right]+{\cal O}(G^3)\ ,\label{IIGhGhXexp}\\
{\bf I}(G_s,G_h,X)&=&-\frac{1}{X}{\bf A}(G){\bf A}(X)\left(1+\frac{4}{d}\frac{G}{X}\right)+{\cal O}(G^3)\ ,\label{IIGsGhXexp}\\
{\bf I}(G_s,G_s,X)&=&\frac{1}{X}\left[{\bf A}(G)\right]^2+{\cal O}(G^3)\ .
\label{IIGsGsXexp}
\eea
Note the factor ${\bf A}(G)$
in ${\bf I}(G_s,G_h,X)$, ${\bf I}(G_s,G_s,X)$, which is a common feature of all ${\bf I}$-functions involving $G_s$, as we will see.

It is convenient to define renormalized versions of the previous results by subtracting subdivergences, as done {\it e.g.}~in \cite{V2,Martin3LV}. Using normal (bold) fonts for the renormalized 
(unrenormalized) functions, as in \cite{Martin3LV}, we have
\be
I(X,Y,Z)=
\lim_{\epsilon\to 0}\left[{\bf I}(X,Y,Z)- {\bf I}_{\rm div}^{(1)}(X,Y,Z)-{\bf I}_{\rm div}^{(2)}(X,Y,Z)\right]\ ,\label{Ir}
\ee
where ${\bf I}_{\rm div}^{(1)}(X,Y,Z)$ are the one-loop subdivergences and ${\bf I}_{\rm div}^{(2)}(X,Y,Z)$ are the two-loop ones.

For the current case we have
\bea
{\bf I}_{\rm div}^{(1)}(G_h,G_h,X)&=&\frac{1}{\epsilon}\left(1-\frac{2G}{X}-\frac{2G^2}{X^2}\right){\bf A}(X)+{\cal O}(G^3)\ ,\\
{\bf I}_{\rm div}^{(2)}(G_h,G_h,X)&=&
\left(\frac{X}{2}-\frac{G^2}{X}\right)\left(
\frac{1}{\epsilon^2}-\frac{1}{\epsilon}\right)-G\left(
\frac{1}{\epsilon^2}+\frac{1}{\epsilon}\right)
+{\cal O}(G^3)\ ,\\
{\bf I}_{\rm div}^{(1)}(G_s,G_h,X)&=&\frac{1}{\epsilon}\left[{\bf A}(G)+\frac{G}{X}{\bf A}(X)\right]\left(1+\frac{G}{X}\right)+{\cal O}(G^3)\ ,\\
{\bf I}_{\rm div}^{(2)}(G_s,G_h,X)&=&
\frac{G}{\epsilon^2}+\frac{G^2}{X}\left(\frac{1}{\epsilon^2}-\frac{1}{2\epsilon}\right)+{\cal O}(G^3)\ ,\\
{\bf I}_{\rm div}^{(1)}(G_s,G_s,X)&=&-\frac{2G{\bf A}(G)}{\epsilon\ X}+{\cal O}(G^3)\ ,\\
{\bf I}_{\rm div}^{(2)}(G_s,G_s,X)&=&
-\frac{G^2}{\epsilon^2X}+{\cal O}(G^3)\ 
\eea
We then get
\bea
I(G_h,G_h,X)&=&\left(1-\frac{2G}{X}-\frac{2G^2}{X^2}\right) I(0,0,X)+4G\left(L_X-\frac32-\frac{G}{X}\right)+{\cal O}(G^3)\ ,\\
I(G_s,G_h,X)&=&-\frac{1}{X}A(G)A(X)+\frac{G^2}{2X}\left(3L_X+3L_G-2L_XL_G-\frac{9}{2}\right)+{\cal O}(G^3)\ ,\\
I(G_s,G_s,X)&=&\frac{1}{X}\left[A(G)\right]^2+{\cal O}(G^3)\ ,
\eea
where 
\be
A(Z)\equiv Z(L_Z-1).
\ee
is the renormalized version of ${\bf A}(Z)$, and
\be
I(0,0,X)=X\left(-\frac52-\frac{\pi^2}{6}+2L_X-\frac12 L_X^2\right)\ .
\ee

For the particular case $X=0$, the previous discussion simply reduces to ${\bf I}(G_s,G_s,0)={\bf I}(G,G,0)$. Similarly, one has ${\bf I}(G_s,0,0)={\bf I}(G,0,0)$.

Working to all orders in $G$, the splitting of the two-loop basis integral ${\bf I}(G,G,X)$ gives
\bea
{\bf I}(G_s,G_s,X) &=&  \frac{\left[{\bf A}(G)\right]^2}{X}\,  _2F_1\left(1,\frac32-\epsilon;3-2\epsilon;\frac{4G}{X}\right) ,\nonumber\\
{\bf I}(G_s,G_h,X) &=&- \frac{{\bf A}(G){\bf A}(X)}{X}\, _2F_1\left(1,\frac12;2-\epsilon;\frac{4G}{X}\right),
\nonumber\\
{\bf I}(G_h,G_h,X) &=&-2 \sqrt{X(X-4G)}\left[\frac{X(X-4G)}{(4\pi\mu^2)^2}\right]^{-\epsilon}\pi\csc(\pi\epsilon)\Gamma(2\epsilon-2)\ .\label{IGGX}
\eea
It is straightforward to check analytically that one has
\be
{\bf I}(G,G,X) = {\bf I}(G_s,G_s,X)+2\, {\bf I}(G_s,G_h,X)+{\bf I}(G_h,G_h,X)\ .
\label{IGGXsum}
\ee
To prove this, let us use as starting point the expression for ${\bf I}(G,G,X)$, valid for generic $d=4-2\epsilon$, as derived in \cite{V2} in terms of the incomplete beta function:
\bea
{\bf I}(G,G,X)&=&-\Gamma(\epsilon)\Gamma(\epsilon-1)4^{-1+\epsilon}\sqrt{X(X-4G)}\left[\frac{X(X-4G)}{(4\pi\mu^2)^2}\right]^{-\epsilon}\nonumber\\
&&\times\left\{2\beta\left(1-\frac{4G}{X};-\frac12+\epsilon,1-\epsilon\right)-\beta\left(\frac{X(X-4G)}{(X-2G)^2};-\frac12+\epsilon,1-\epsilon\right)\right\} .\label{IGGXV2}
\eea
One then rewrites this expression in terms of hypergeometric functions using $B(z;p,q)=(z^p/p)\, _2F_1(p,1-q;p+1;z)$.
Next, the resulting hypergeometric functions can be transformed using the quadratic transformation
\be
 _2F_1\left(a,a+\frac12;c;z\right) =\left(\frac{1+\sqrt{1-z}}{2}\right)^{-2a}\, 
_2F_1\left(2a,2a-c+1;c,\frac{1-\sqrt{1-z}}{1+\sqrt{1-z}}\right)\, ,\label{t1}
\ee
as given by \cite{AS} [formula (15.3.19)] , and 
 the identity
\bea
_2F_1(a,b;c;z)&=&
(1-z)^{c-a-b}\frac{\Gamma(c)\Gamma(a+b-c)}{\Gamma(a)\Gamma(b)}\, _2F_1(c-a,c-b;c-a-b+1;1-z)
\nonumber\\
&&+\frac{\Gamma(c)\Gamma(c-a-b)}{\Gamma(c-a)\Gamma(c-b)}\, _2F_1(a,b;a+b-c+1;1-z)\ ,
\label{t2}
\eea
 [formula (15.3.6) of \cite{AS}]. In this way one is able to recast (\ref{IGGXV2}) as a sum of three terms that correspond precisely to the split terms of  (\ref{IGGXsum}) given in (\ref{IGGX}).
As a cross-check, one can show that the expansion of the full results of Eqs.~(\ref{IGGX}) in powers of $\epsilon$ and $G$ reproduces
the previous expanded results of Eqs.~(\ref{IIGhGhXexp}-\ref{IIGsGsXexp}). 

As a final side comment, note that the split functions given in Eq.~(\ref{IGGX}) have poles at lower dimensions, corresponding to poles in $\epsilon=1,2$ which are closely related to each other in such a way that the leading divergence cancels in their sum, Eq.~(\ref{IGGXsum}). As an example, for $d=2$ (or $\epsilon=1$)
one has
\bea
{\bf I}(G_s,G_s,X)=-{\bf I}(G_s,G_h,X)={\bf I}(G_h,G_h,X)=\frac{\mu^4[\Gamma(\epsilon-1)]^2}{\kappa\sqrt{X(X-4G)}} +{\cal O}\left(\frac{1}{\epsilon-1}\right)\ .
\eea
This fact is generic and holds for other splittings we derive in the next subsections.

\section{Splitting of \bma{{\bf I}(G,X,0)}}

The splitting of ${\bf I}(G,X,0)$ into ${\bf I}(G_s,X,0)$ and
${\bf I}(G_h,X,0)$, up to ${\cal O}(G^2)$, gives 
\bea
{\bf I}(G_h,X,0)&=&{\bf I}(0,0,X)\left[1-(d-3)\frac{G}{X}+(d-3)(d-4)\frac{G^2}{2X^2}\right]+{\cal O}(G^3)\ ,\label{IIGhX0exp}\\
{\bf I}(G_s,X,0)&=&-\frac{1}{X}{\bf A}(G){\bf A}(X)\left[1-\frac{(d-4)}{d }\frac{G}{X}\right]+{\cal O}(G^3)\ .
\label{IIGsX0exp}
\eea
Note again the factor ${\bf A}(G)$
in ${\bf I}(G_s,X,0)$, as anticipated.

As for the previous case, the renormalized functions are defined by
Eq.~(\ref{Ir}) with
\bea
{\bf I}_{\rm div}^{(1)}(G_h,X,0)&=&\frac{1}{\epsilon}\left(1-\frac{G}{X}\right){\bf A}(X)+{\cal O}(G^3)\ ,\\
{\bf I}_{\rm div}^{(2)}(G_h,X,0)&=&
\frac{1}{2\epsilon^2}(X-G)-\frac{1}{2\epsilon}(X+G)+\frac{G^2}{2\epsilon X}+{\cal O}(G^3)\ ,\\
{\bf I}_{\rm div}^{(1)}(G_s,X,0)&=&\frac{1}{\epsilon}\left[{\bf A}(G)+\frac{G}{X}{\bf A}(X)\right]+{\cal O}(G^3)\ ,\\
{\bf I}_{\rm div}^{(2)}(G_s,X,0)&=&
\frac{G}{\epsilon^2}-\frac{G^2}{2\epsilon X}+{\cal O}(G^3)\ .
\eea
We then get
\bea
I(G_h,X,0)&=&\left(1-\frac{G}{X}\right) I(0,0,X)+2G
\left(1-\frac{G}{2X}\right)L_X-3G+\frac{G^2}{2X}+{\cal O}(G^3)\ ,\\
I(G_s,X,0)&=&-\frac{1}{X}A(G)A(X)+\frac{G^2}{4X}\left(2L_G+2L_X-5\right)+{\cal O}(G^3)\ .
\eea

Working to all orders in $G$, the splitting of ${\bf I}(G,X,0)$ gives
\bea
{\bf I}(G_s,X,0)&=&-\frac{{\bf A}(G){\bf A}(X)}{X} \,  _2F_1\left(1,\epsilon;2-\epsilon;\frac{G}{X}\right) ,\\
{\bf I}(G_h,X,0)&=&-2 (X-G)\left[\frac{X-G}{4\pi\mu^2}\right]^{-2\epsilon}\pi\csc(\pi\epsilon)\Gamma(2\epsilon-2)\ .
\eea
As a cross-check, one can show that the expansion of these full results in powers of $\epsilon$ and $G$ reproduces
the previous expanded results of Eqs.~(\ref{IIGhX0exp},\ref{IIGsX0exp}). 
It is also straightforward to check that 
\be
{\bf I}(G,X,0)={\bf I}(G_s,X,0)+{\bf I}(G_h,X,0)\ .
\label{IGX0sum}
\ee
In order to do so, one can start from the general expression as derived in \cite{V2} in terms of the incomplete beta function:
\bea
{\bf I}(G,X,0)&=&-
\frac{1}{4^{1-\epsilon}}(X-G)\left[\frac{X-G}{4\pi^2\mu^2}\right]^{-2\epsilon}\Gamma(\epsilon-1)\Gamma(\epsilon)\,\beta\left(\frac{(X-G)^2}{(X+G)^2};-\frac{1}{2}+\epsilon,1-\epsilon\right)
\nonumber\\
&=& X\left[\frac{X}{4\pi^2\mu^2}\right]^{-2\epsilon}
\frac{\Gamma(\epsilon-1)\Gamma(\epsilon)}{(1-2\epsilon)}\, _2F_1\left(-1+2\epsilon,\epsilon;2\epsilon;1-\frac{G}{X}\right)\ .
\eea
rewritten in the last line in terms of the hypergeometric function
and then using (\ref{t2}) to reduce the expression to agree with  (\ref{IGX0sum}).
 
\section{Splitting of \bma{{\bf I}(G,X,X)}}

As for previous cases, the ${\cal O}(G^2)$  splitting of ${\bf I}(G,X,X)$ into  ${\bf I}(G_s,X,X)$ and ${\bf I}(G_h,X,X)$ can be obtained in a direct way and one gets 
\bea
{\bf I}(G_h,X,X)&=&{\bf I}(0,X,X)\left[1+\frac{(d-3)}{2(d-5)}\frac{G}{X}+\frac{(d-3)(d-4)}{4(d-5)(d-7)}\frac{G^2}{X^2}\right]+{\cal O}(G^2)\ ,\label{IIGhXXexp}\\
{\bf I}(G_s,X,X)&=&-{\bf A}(G){\bf A}'(X)\left[1-\frac{(d-4)}{12}\frac{G}{X}\right]+{\cal O}(G^3)\ ,
\label{IIGsXXexp}
\eea
where ${\bf A}'(X)\equiv d{\bf A}(X)/dX=(d-2){\bf A}(X)/(2X)$. Note again the factor ${\bf A}(G)$
in ${\bf I}(G_s,X,X)$, as anticipated.

As for previous cases, the renormalized functions are defined by
Eq.~(\ref{Ir}) with
\bea
{\bf I}_{\rm div}^{(1)}(G_h,X,X)&=&\frac{1}{\epsilon}\left[2{\bf A}(X)-G {\bf A}'(X)\right]+{\cal O}(G^3)\ ,\\
{\bf I}_{\rm div}^{(2)}(G_h,X,X)&=&
\frac{1}{2\epsilon^2}(2X-G)-\frac{1}{2\epsilon}(2X+G)+\frac{G^2}{6\epsilon X}+{\cal O}(G^3)\ ,\\
{\bf I}_{\rm div}^{(1)}(G_s,X,X)&=&\frac{1}{\epsilon}\left[{\bf A}(G)+G{\bf A}'(X)\right]+{\cal O}(G^3)\ ,\\
{\bf I}_{\rm div}^{(2)}(G_s,X,X)&=&
\frac{G}{\epsilon^2}-\frac{G^2}{6\epsilon X}+{\cal O}(G^3)\ .
\eea
We then get the renormalized results
\bea
I(G_h,X,X)&=&\left(1-\frac{G}{2X}\right) I(0,X,X)+G\left(3L_X-1\right)-\frac{G^2}{3X}\left(L_X+\frac56\right)+{\cal O}(G^3)\ ,\\
I(G_s,X,X)&=&-A(G)A'(X)+\frac{G^2}{6X}\left(L_G+L_X-1\right)+{\cal O}(G^3)\ ,
\eea
where
\be
I(0,X,X)=X(-5+4L_X-L_X^2)\ .
\ee

 The splitting of ${\bf I}(G,X,X)$ to all orders in $G$ gives
 \bea
{\bf I}(G_s,X,X)&=& -{\bf A}(G){\bf A}'(X)
\, _2F_1\left(1,\epsilon;\frac{3}{2};\frac{G}{4X}\right) ,\label{IGsXX}\\
{\bf I}(G_h,X,X)&=&-\frac{{\bf A}(X){\bf A}'(X)}{(1-2\epsilon)}~
_2F_1\left(1,-1+2\epsilon;\frac12+\epsilon;\frac{G}{4X}\right) .
\eea
It is straightforward to check that 
\be
{\bf I}(G,X,X)={\bf I}(G_s,X,X)+{\bf I}(G_h,X,X)\ .
\ee
Again, one can take as starting point the general expression for ${\bf I}(G,X,X)$ as obtained in \cite{V2} 
\bea
{\bf I}(G,X,X)&=&-2\sqrt{G(4X-G)}\left[\frac{G(4X-G)}{(4\pi^2\mu^2)^2}\right]^{-\epsilon}\pi \Gamma(2\epsilon-2)\nonumber\\
&&+\frac{1}{2}\left[\frac{X}{4\pi^2\mu^2}\right]^{-\epsilon}\Gamma(\epsilon)\Gamma(\epsilon-1)\left\{2G\left[\frac{G}{4\pi^2\mu^2}\right]^{-\epsilon}\, _2F_1\left(1,\epsilon;\frac32;\frac{G}{4X}\right)\right.
\nonumber\\
&&\left.+
(2X-G)\left[\frac{X}{4\pi^2\mu^2}\right]^{-\epsilon}\, _2F_1\left(1,\epsilon;\frac32;\frac{(2X-G)^2}{4X^2}\right)\right\}
\eea
and reduce it to a suitable form transforming the hypergeometric functions involved, using in particular the quadratic transformation
[given by \cite{Goursat}, eq.~(27) of page 118]
\bea
&&\frac{4\sqrt{\pi}\Gamma(a+b-1/2)}{\Gamma(a-1/2)\Gamma(b-1/2)}\sqrt{z}\, _2F_1\left(a,b;\frac32;z\right) =\,\label{t3} \\
&&
_2F_1\left(2a-1,2b-1;a+b-\frac12,\frac{1+\sqrt{z}}{2}\right)-\,_2F_1\left(2a-1,2b-1;a+b-\frac12,\frac{1-\sqrt{z}}{2}\right)\, .\nonumber
\eea

\section{Splitting of \bma{{\bf I}(G,X,Y)}}

The splitting of ${\bf I}(G,X,Y)$ into ${\bf I}(G_s,X,Y)$ and ${\bf I}(G_h,X,Y)$, up to ${\cal O}(G^2)$, gives 
\bea
{\bf I}(G_h,X,Y)&=&{\bf I}(0,X,Y)\left\{1-(d-3)\frac{G(X+Y)}{(X-Y)^2}\right.\nonumber\\
&+&\left.(d-3)\left[(d-4)(X^2+Y^2)+2(d-6)XY\right]\frac{G^2}{2(X-Y)^4}
\right\}\nonumber\\
&+&(d-2)G\left[-1+(d-5)\frac{G(X+Y)}{2(X-Y)^2}\right]\frac{{\bf A}(X){\bf A}(Y)}{(X-Y)^2}+{\cal O}(G^3),\\
{\bf I}(G_s,X,Y)&=&-{\bf A}(G)\frac{{\bf A}(X)-{\bf A}(Y)}{X-Y}+\left.\frac{G {\bf A}(G)}{(X-Y)^3}\right\{X {\bf A}(Y)-Y {\bf A}(X)\nonumber\\
&+&\left.\frac{(d-4)}{d}\left[X {\bf A}(X)-Y {\bf A}(Y)\right]\right\}+{\cal O}(G^3)\ .
\eea
Note again the factor ${\bf A}(G)$
in ${\bf I}(G_s,X,Y)$, as expected.

As for previous cases, the renormalized functions are defined by
Eq.~(\ref{Ir}) with
\bea
{\bf I}_{\rm div}^{(1)}(G_h,X,Y)&=&\frac{1}{\epsilon}{\bf A}(X)
\left[1-\frac{G}{X-Y}-\frac{G^2Y}{(X-Y)^3}
\right]\nonumber\\
&+&\frac{1}{\epsilon}{\bf A}(Y)
\left[1+\frac{G}{X-Y}+\frac{G^2X}{(X-Y)^3}
\right]+{\cal O}(G^3)\ ,\\
{\bf I}_{\rm div}^{(2)}(G_h,X,Y)&=&
\frac{1}{2\epsilon^2}(X+Y-G)-\frac{1}{2\epsilon}(X+Y+G)+\frac{1}{2\epsilon}\frac{G^2(X+Y)}{(X-Y)^2}+{\cal O}(G^3)\ ,\\
{\bf I}_{\rm div}^{(1)}(G_s,X,Y)&=&\frac{1}{\epsilon}\left[{\bf A}(G)+G\ \frac{{\bf A}(X)-{\bf A}(Y)}{X-Y}\right]-\frac{G^2\left[X{\bf A}(Y)-Y{\bf A}(X)\right]}{\epsilon (X-Y)^3}+{\cal O}(G^3),\\
{\bf I}_{\rm div}^{(2)}(G_s,X,Y)&=&
\frac{G}{\epsilon^2}-\frac{G^2(X+Y)}{2\epsilon (X-Y)^2}+{\cal O}(G^3)\ .
\eea
We then get the renormalized results
\bea
I(G_h,X,Y)&=&\left[1-\frac{G(X+Y)}{(X-Y)^2}-\frac{2G^2XY}{(X-Y)^4}\right] I(0,X,Y) \nonumber\\
&-&
\frac{G}{(X-Y)^2}
\left\{(X+Y)^2-2\left[X A(X)+Y A(Y)-A(X)A(Y)\right]\right\}\nonumber\\
&-&\left.\frac{G^2}{2(X-Y)^4}\right\{(X+Y)^3+2(X-Y)\left[X A(X)-Y A(Y)\right]\nonumber\\
&+&\left.2(X+Y)A(X)A(Y)-4XY\left[A(X)+A(Y)\right]\frac{}{}\right\}+{\cal O}(G^3)\ ,\\
I(G_s,X,Y)&=&-A(G)\frac{A(X)- A(Y)}{X-Y}+\frac{G A(G)}{(X-Y)^3}\left[XA(Y)-Y A(X)\right]\nonumber\\
&+&\frac{G^2}{4(X-Y)^3}\left[(2L_X+2L_G-5)X^2-(2L_Y+2L_G-5)Y^2\right]+{\cal O}(G^3)\ ,
\eea
with
\bea
I(0,X,Y)&=&\frac14 (X+Y)\left[(L_X-L_Y)^2-10\right]-\frac12 \left[X L_X(L_X-4)+ Y L_Y(L_Y-4)\right]\nonumber\\
&+&\frac12(X-Y)\left[{\rm Li}_2(1-X/Y)-{\rm Li}_2(1-Y/X)\right]\ .
\eea

Working to all orders in $G$, the splitting of ${\bf I}(G,X,Y)$ gives
 \bea
{\bf I}(G_s,X,Y)&=&{\bf A}(G)\left[\frac{Y}{4\pi\mu^2}\right]^{-\epsilon}\,\nonumber\\
&&\times
\sum_{n=0}^{\infty}
\left(\frac{G}{Y}\right)^n\frac{\Gamma(n+1)\Gamma(n+\epsilon)}{\Gamma(2n+2)}
\, _2F_1\left(n+1,n+\epsilon;2n+2;1-\frac{X}{Y}\right)  ,\\
{\bf I}(G_h,X,Y)&=&-{\bf A}(Y)\left[\frac{Y}{4\pi\mu^2}\right]^{-\epsilon}\,\label{IGsXY}\nonumber\\
&&\times
\sum_{n=0}^{\infty}
\left(\frac{G}{Y}\right)^n\frac{\Gamma(n+\epsilon)\Gamma(n-1+2\epsilon)}{\Gamma(2n+2\epsilon)}\, _2F_1\left(n+\epsilon,n-1+2\epsilon;2n+2\epsilon;1-\frac{X}{Y}\right) ,\nonumber\\
&&\label{IGhXY}
\eea
and it can be readily checked that these expressions reproduce  ${\bf I}(G_s,X,X)$ and ${\bf I}(G_h,X,X)$ in the limit $Y\rightarrow X$. Notice also that these expressions are
symmetric in $X,Y$ as can be verified by making use of the Pfaff transformation $\, _2F_1(a,b;c;z)=(1-z)^{-b}\,_2F_1(c-a,b;c;z/(z-1))$.
 
One can also show that the sum of split parts reproduces the full result
\be
{\bf I}(G,X,Y)={\bf I}(G_s,X,Y) +{\bf I}(G_h,X,Y)\ .
\ee 
As before, take as starting point the expression for ${\bf I}(G,X,Y)$ as derived in \cite{V2}, which can be rewritten as:
\bea
{\bf I}(G,X,Y)&=&-\frac{1}{2}\Gamma(\epsilon)\Gamma(\epsilon-1) \left\{a_{GXY} \left[\frac{a_{GXY}}{4\pi\mu^2}\right]^{-2\epsilon}\beta\left(\frac{4a^2_{GXY}}{(X+Y-G)^2};-\frac12+\epsilon,1-\epsilon\right)
\right.\nonumber\\
&&-(G+X-Y)\left[\frac{GX}{(4\pi\mu^2)^2}\right]^{-\epsilon}
\, _2F_1\left(1,\epsilon;\frac32;\frac{(G+X-Y)^2}{4GX}\right)
\nonumber\\
&&\left.
-(G+Y-X)\left[\frac{GY}{(4\pi\mu^2)^2}\right]^{-\epsilon}
\, _2F_1\left(1,\epsilon;\frac32;\frac{(G+Y-X)^2}{4GX}\right)
\right\} \, ,\label{IGXYV2}
\eea
with
\be
a^2_{GXY}\equiv \frac{1}{4}\left[(X-Y)^2+G^2-2G(X+Y)\right]\ .
\ee
In fact, the term of Eq.~(\ref{IGXYV2}) corresponding to the  incomplete beta function gives  ${\bf I}(G_h,X,Y)$ and the last two terms give ${\bf I}(G_s,X,Y)$. To show how ${\bf I}(G_h,X,Y)$ is reproduced, rewrite
the incomplete beta function in terms of a hypergeometric function as done after Eq.~(\ref{IGGXV2}) and use on it the quadratic transformation of Eq.~(\ref{t1}). After a Pfaffian transformation one gets
\be
{\bf I}(G_h,X,Y)= \sqrt{X Y} \left[\frac{\sqrt{XY}}{4\pi\mu^2}\right]^{-2\epsilon}\frac{\Gamma(\epsilon)\Gamma(\epsilon-1)}{(1-2\epsilon)}\, _2F_1\left(1,2\epsilon-1;\frac12+\epsilon;\frac{(\sqrt{X}-\sqrt{Y})^2-G}{-4\sqrt{X Y}}\right) ,
\label{IGhXYs}
\ee
One can then show that this agrees with the series in Eq.~(\ref{IGhXY}) by expanding the hypergeometric function
above in powers of $G$ and comparing terms of order $G^n$ making use of the identity (46) in \cite{Goursat} (p.120)
\be
_2F_1(\alpha,\beta;2\beta;x) = (1-x)^{-\alpha/2}\, _2F_1\left(\alpha,2\beta-\alpha;\beta+\frac12;\frac{(1-\sqrt{1-x})^2}{-4\sqrt{1-x}}\right)\ .
\ee
Finally the expression 
\bea
{\bf I}(G_s,X,Y)&=&
\frac{1}{2}\Gamma(\epsilon)\Gamma(\epsilon-1) \left\{(G+X-Y)\left[\frac{GX}{(4\pi\mu^2)^2}\right]^{-\epsilon}
\, _2F_1\left(1,\epsilon;\frac32;\frac{(G+X-Y)^2}{4GX}\right)\right.
\nonumber\\
&&\left.
+(G+Y-X)\left[\frac{GY}{(4\pi\mu^2)^2}\right]^{-\epsilon}
\, _2F_1\left(1,\epsilon;\frac32;\frac{(G+Y-X)^2}{4GY}\right)
\right\} \, ,\label{IGsXYV2}
\eea
can be shown to reproduce ${\bf I}(G_s,X,X)$ in Eq.~(\ref{IGsXX}) when $X=Y$ and by expanding in powers of $G/X$, $G/Y$ and $1-X/Y$
it is not difficult to reproduce the infinite series expression we obtained in Eq.~(\ref{IGsXY}).
 
 The expansion of ${\bf I}(G_h,X,Y)$ in $\epsilon$ is
\bea
{\bf I}(G_h,X,Y)&=&\frac{1}{2}(X+Y-G) \left\{-\frac{1}{\epsilon^2}+\frac{1}{\epsilon}\left(L_X+L_Y-3-z\ln r\right)
\right.\nonumber\\
&&+z\left[\left(L_X+L_Y-3+2\ln(1-r)-\frac12 \ln r\right)\ln r-\frac{\pi^2}{3}+2\mathrm{Li}_2(r)\right]\nonumber\\
&&\left.-\frac12\left[(L_X+L_Y-3)^2+5+\frac{\pi^2}{3}\right]\right\}+{\cal O}(\epsilon)\ ,
\eea
where we have used
\be
z\equiv \frac{2a_{GXY}}{X+Y-G}\, ,\;\; 
r\equiv \frac{1-z}{1+z}
\ .
\ee
To derive the previous expression we first transformed the hypergeometric function in (\ref{IGhXYs}) using identity (40) in \cite{Goursat}, p.120, 
\be
_2F_1\left(\alpha,\beta;\frac{\alpha+\beta+1}{2};z\right)=
\left(\sqrt{1-z}+\sqrt{-z}\right)^{-2\alpha}\, _2F_1\left(\alpha,\frac{\alpha+\beta}{2};\alpha+\beta;\frac{4\sqrt{z(z-1)}}{(\sqrt{1-z}+\sqrt{-z})^2}\right)\ ,
\label{goursat40}
\ee
and then performed the $\epsilon$-expansion
\be
_2F_1(1,\epsilon;2\epsilon;x)=\frac{1}{2(1-x)}\left\{
2-x+x\epsilon \ln(1-x)+x\epsilon^2\left[\frac12\ln^2(1-x)+2\mathrm{Li}_2(x)\right]+{\cal O}(\epsilon^3)\right\}\ ,
\label{2F1exp}
\ee 
using the techniques of \cite{MUW}.
Expanding further in $G$ we arrive at
\bea
{\bf I}(G_h,X,Y)&=&\frac{1}{2} \left\{-\frac{1}{\epsilon^2}(X+Y-G)\right.\nonumber\\
&& +\frac{1}{\epsilon}\left[X(2L_X-3)+Y(2L_Y-3)-G\left(2\frac{XL_X-YL_Y}{X-Y}-1\right)\right]\nonumber\\
&&-\frac12(X+Y)\left[(L_X+L_Y-3)^2+5+\frac{\pi^2}{3}\right]+
\frac{G}{2}\left[(L_X+L_Y+1)^2+5+\frac{\pi^2}{3}\right]\nonumber\\
&&+\left(X-Y-G\frac{X+Y}{X-Y}\right)\left[\ln\frac{X}{Y}\left[3-\frac{3L_Y+L_X}{2}-2\ln\left(1-\frac{X}{Y}\right)\right]+\frac{\pi^2}{3}\right.\nonumber\\
&&\left.-2\mathrm{Li}_2\left(\frac{X}{Y}\right)\right]+\left. 2G\left(\frac{X+Y}{X-Y}\ln\frac{X}{Y}-L_X-L_Y-1\right)\right\}+{\cal O}(\epsilon)+{\cal O}(G^2),\nonumber\\
&&\label{IIGhXY}
\eea

We also get the following $\epsilon$ expansion
\bea
_2F_1\left(1,\epsilon;\frac32;x\right)&=&1+2\epsilon\left(1-\frac{\sqrt{1-x}}{\sqrt{x}}\theta\right)\nonumber\\
&+&2\epsilon^2\left\{2+\frac{\sqrt{1-x}}{\sqrt{x}}\left[\left(\ln(4-4x)-2\right)\theta-{\rm Cl}_2(\pi-2\theta)\right]\right\}+{\cal O}(\epsilon^3) ,
\eea
with $\theta=\arcsin\sqrt{x}$ and ${\rm Cl}_2$ is the Clausen function.

\section{Result for $\bma{ \Delta_2}$}
In this appendix we present the two-loop result for the shift of the Goldstone mass (used in resumming IR divergences) calculated from (\ref{Delta2}) to order ${\cal O}(G^0)$:
\bea
\Delta_2 &=& \left.\frac{3\lambda  (g^2+g_Z^2)^3}{2g_Z^4}I(0,W,Z)- 6 \lambda ^2 I(0,0,H)- 15  \lambda ^2I(H,H,H)
-12 y_t^2 \lambda I(0,0,T)\right.\nonumber\\
&+& \frac32 g^2 \left[\frac{\left(2 g^4+8 g'{}^2 g^2+5 g'{}^4\right) \lambda }{g_Z^4}-6 g^2\right] I(0,0,W)-3 \left(g^2-2 y_t^2\right) \left(y_t^2+g^2\right)I(0,W,T) \nonumber\\
&+& \frac{1}{12} \left[\frac{18 (g^2-2 g'{}^2) \lambda  g^2}{g_Z^2}-63 g^4-6 g'{}^2 g^2-103 g'{}^4\right] I(0,0,Z)\nonumber\\
&+&
\frac{1}{8g_Z^2} (4 g^2-g_Z^2) \left(12 g^4+20 g_Z^2 g^2+g_Z^4\right) I(W,W,Z)+\frac92  \left(3 y_t^2-4  \lambda \right)y_t^2I(H,T,T) \nonumber\\
&+&\frac{1}{12}\left[-9 g^4+6 g'{}^2 g^2-17 g'{}^4+\frac{2 \left(9 g^4+66 g'{}^2 g^2-7 g'{}^4\right) y_t^2}{g_Z^2}\right] I(Z,T,T)
\nonumber\\
&+&
\frac13  y_t^2\left[6\left(8g_3^2-6y_t^2+3\lambda\right)
-27 g_Z^2-26g'{}^2+64\frac{g'{}^4}{g_Z^2}\right]A(T)
\nonumber\\
&+&
\frac{1}{48}  \left[-444 g^2 \lambda +132 g'{}^2 \lambda -\frac{32 A(T) \left(9 g^4-6 g'{}^2 g^2+17 g'{}^4\right)}{g_Z^2 \phi ^2}\right.\nonumber\\
&+&\left.
\frac{1}{g_Z^2}\left[317 g^6+221 g'{}^2 g^4+647 g'{}^4 g^2+455 g'{}^6-4 \left(63 g^4+30 g'{}^2 g^2+95 g'{}^4\right) y_t^2\right]\right]A(Z)
\nonumber\\
&+&
 \left\{\frac{1}{24} g^2 \left[605 g^2-252 y_t^2+39 g'{}^2-\frac{288 g^4+12 (13 g^2+7 g'{}^2) \lambda }{g_Z^2}\right]+\frac{12  \left(y_t^2-g^2\right)}{\phi ^2}A(T)\right.
\nonumber\\
&+&\left.
\frac{1}{g_Z^4 \phi ^2} \left[15 g^6+5 g'{}^2 g^4-11 g'{}^4 g^2-g'{}^6+6 \left(8 g^4+8 g'{}^2 g^2+g'{}^4\right) \lambda \right]
A(Z)\right\} A(W)
\nonumber\\
&+&
\frac{3}{2} A(H) \left[-7 y_t^4+22 \lambda^2
+6 y_t^2\left(\lambda -\frac{2 A(T)}{\phi ^2}\right)\right]+\frac{A(W)^2}{2\phi^2} \left(48\frac{g^4}{g_Z^2}+g_Z^2-34g^2\right)
\nonumber\\
&+&
\frac{A(T)^2}{3g_Z^2 \phi ^2} \left[9 g_Z^2\left(5 y_t^2+32 g_3^2+2 \lambda \right) +9 g^4+g'{}^2 \left(90g^2+17g'{}^2
  \right)\right]+
 \frac{9 A(H)^2 \lambda }{2\phi ^2}
 \nonumber\\
&+&\left\{-
y_t^2 \left[\frac{1}{48} \left(310 g^4-320 g^2 g_Z^2+91 g_Z^4\right)+18 \lambda ^2+\frac{y_t^2}{6}  \left(64 \frac{g^4}{g_Z^2}-77 g^2 - 5 g_Z^2-36 \lambda \right)\right.\right.
  \nonumber\\
&-&
 24 g_3^2 y_t^2-9 y_t^4\left]+
\frac38 (\lambda+8g^2)\frac{g^6}{g_Z^2}+
\frac{1}{192}\left(872 g^6- 496g^4g_Z^2+718 g^2 g_Z^4-497 g_Z^6\right)\right.\nonumber\\
&+&
\left.
\frac{\lambda}{16}
 \left[92g^4+12g^2g_Z^2+7g_Z^4+32(2g^2+g_Z^2)\lambda-960\lambda^2\right]\right\}\phi^2
\nonumber\\
&+&
\left\{\frac32 \lambda  (g^2+8 \lambda )I(0,W,H) +
\frac{1}{16}  \left(\frac{3 g^6}{g^2-2 \lambda }-30 g^4+88 \lambda  g^2-224 \lambda ^2\right)I(W,W,H)\right.
\nonumber\\
&+&\frac{1}{4(g^2-2\lambda)\phi^2}\left[\frac{9g^8\phi^4}{8}-\frac{3g^6\phi^2}{4}\left[2A(W)+A(H)\right]
+2\left(9 g^4-31 \lambda  g^2+32 \lambda ^2\right) A(W)A(H)
\right.
\nonumber\\
&-&\left.
\left(7 g^4-36 \lambda  g^2+56 \lambda ^2\right)A(W)^2\left] +
 \frac{g^2}{8}(27g^2-28\lambda)A(H)+
\frac12\left[g\rightarrow g_Z, W\rightarrow Z\right]\right\}\right.
 .
\eea

\end{document}